\documentclass[fleqn,usenatbib]{mnras}

\providecommand{\sorthelp}[1]{}
 
\usepackage[utf8]{inputenc}
\usepackage{csquotes}
\usepackage{bm}
\usepackage{amsmath}
\usepackage{mathtools} 
\usepackage{amssymb}
\usepackage[capitalise]{cleveref}
\usepackage{graphicx}% Include figure files
\usepackage[usenames,dvipsnames]{color}
\usepackage{hyperref}
\usepackage{widetext}
\hypersetup{
    colorlinks=true,
    linkcolor=blue,
    filecolor=magenta,      
    urlcolor=cyan,
    pdftitle={PNG and kNN-CDFs},
    pdfpagemode=FullScreen,
    }
\let\oldhat\hat
\renewcommand{\hat}[1]{\oldhat{\mathbf{#1}}}

\usepackage{bm}

\title[]{Small-scale signatures of primordial non-Gaussianity in k-Nearest Neighbour cumulative distribution functions}

\author[]{William R. Coulton,$^{1}$  Tom Abel,$^{2,3,4}$  and Arka Banerjee$^{5}$ \\
$^{1}$ Center for Computational Astrophysics, Flatiron Institute, 162 5th Avenue, New York, NY 10010, USA \\
$^{2}$ Kavli Institute for Particle Astrophysics and Cosmology, Stanford University, 452 Lomita Mall, Stanford, CA 94305, USA \\
$^{3}$ Department of Physics, Stanford University, 382 Via Pueblo Mall, Stanford, CA 94305, USA \\
$^{4}$ SLAC National Accelerator Laboratory, 2575 Sand Hill Road, Menlo Park, CA 94025, USA\\
$^{5}$ Department of Physics, Indian Institute of Science Education and Research,
Homi Bhabha Road, Pashan, Pune 411008, India \\
}
\date{\today}

\begin{document}

\maketitle

\begin{abstract}
Searches for primordial non-Gaussianity in cosmological perturbations are a key means of revealing novel primordial physics. However, robustly extracting signatures of primordial non-Gaussianity from non-linear scales of the late-time Universe is an open problem. In this paper,  we apply k-Nearest Neighbor cumulative distribution functions, kNN-CDFs, to the \textsc{quijote-png} simulations to explore the sensitivity of kNN-CDFs to primordial non-Gaussianity. An interesting result is that for halo samples with $M_h<10^{14}$ M$_\odot$/h, the kNN-CDFs respond to \textit{equilateral} PNG in a manner distinct from the other parameters. This persists in the galaxy catalogs in redshift space and can be differentiated from the impact of galaxy modelling, at least within the halo occupation distribution (HOD) framework considered here. kNN-CDFs are related to counts-in-cells and, through mapping a subset of the kNN-CDF measurements into the count-in-cells picture, we show that our results can be modeled analytically. A caveat of the analysis is that we only consider the HOD framework, including assembly bias. It will be interesting to validate these results with other techniques for modeling the galaxy--halo connection, e.g., (hybrid) effective field theory or semi-analytical methods. 
\end{abstract}

\section{Introduction} \label{sec:introduction}
Many theories of the early Universe predict that the statistical distribution of primordial potential perturbations is close to Gaussian, but with small deviations \citep[see e.g.,][for recent reviews]{Chen_2010,Achucarro_2022,Meerburg_2019}. The structure of these deviations, known as primordial non-Gaussianity (PNG), encodes the details of the physical processes governing the evolution of the Universe at that epoch. Primordial properties ranging from the number of particles present, the masses and spins of these particles, the strength of interactions, and primordial symmetries all leave distinct non-Gaussian signatures \citep[e.g.][]{Maldacena_2003,Creminelli_2004,Alishahiha_2004,Chen_2007,Meerburg_2009,Arkani-Hamed_2015,Cabass_2023c}. Thus, characterizing the statistical distribution of primordial perturbations is a powerful way to reveal new information on the early Universel and probe energy scales far beyond the reach of terrestrial experiments.  
In this work, we focus on three templates of non-Gaussianity that are most relevant for Large Scale Structure (LSS) - \textit{local}, \textit{equilateral}, and \textit{orthogonal} \citep{Komatsu_2001,Senatore_2010}. Each of these generates a unique signature in the primordial bispectrum, which characterize the skewness of the primordial perturbations as a function of scale.

To date, observational studies of primordial non-Gaussianity have been driven by measurements of the bispectrum of the cosmic microwave background (CMB) anisotropies \citep[e.g.,][]{komatsu2003,planck2016-l09} and the large-scale distribution of galaxies \citep{DAmico_2022,DAmico_2022b,Cabass_2022a,Cabass_2022b,Cabass_2023b}. Whilst no signatures of primordial non-Gaussianity have yet been detected, large regions of primordial model space have already been ruled out.  New experiments, such as the Dark Energy Spectroscopic Instrument, SPHEREX, the Simons Observatory, and CMB-S4, will provide dramatically expanded and more precise data sets that will significantly improve upon current bounds \citep{dore_2013,DESI_2016,S4_2016,SO_2019}. However, these constraints have not yet reached the regions of greatest theoretical interest that divide qualitatively different regions, such as strong and weakly coupled physics \citep{Cabass_2023a}.

The bispectrum has been used extensively for two reasons: first, it is the optimal statistic to constrain \textit{local}, \textit{equilateral} and \textit{orthogonal} non-Gaussianity in the CMB and in the very large-scale distribution of galaxies \citep{Babich_2005,Philcox_2021}. Second, analytical tools have been developed that can accurately model these observations \citep[e.g.][with the latter for a review]{Baumann_2012,Carrasco_2012,Cabass_2023d}.  However, for measurements of the small-scale distribution of galaxies, where the signal-to-noise ratio (SNR) is high, and where the relation to the primordial anisotropies is non-linear, these statements break down. The non-linear evolution redistributes information from the primordial bispectrum to not only the late-time bispectrum, but also to the trispectrum, pentaspectrum, and beyond (where the trispectrum and pentaspectrum  are the kurtosis and 5th moment as a function scale). This means that analyses based purely on the late-time bisepctrum are not accessing all of the available information. One approach is to include these higher-order correlation functions in the analysis, however it is highly challenging to compute these statistics \citep{Philcox_2021a}. Further, the ability to model the bispectrum relies on a perturbative analysis which is typically only valid on large scales. A final challenge of this approach is that the non-linear processes governing structure evolution generate late-time non-Gaussianities, even in the absence of PNG. These non-Gaussianities can mimic the bispectrum signatures of PNG and thereby bias inferences. When removing these biases, we need to marginalize over the uncertainties in our understanding of these processes, for example in how galaxies occupy halos. This significantly degrades the resulting PNG constraints, especially for \textit{equilateral} non-Gaussianity \citep{Baldauf_2016,Lazanu_2017,Baumann_2022, Cabass_2023a}.

In this work we explore the efficacy of k-nearest neighbour cumulative distributions functions (kNN-CDFs), an alternative summary statistic to the hierarchy of $N$-point correlations, to constrain PNG. kNN-CDFs describe the volume-averaged probability of finding at least k objects, in our case dark matter halos or galaxies, within a sphere of radius $R$. Recent work  \citep{Banerjee_2021,Banerjee_Abel_cross,kNN_SDSS,Banerjee_Abel_TFcross}  have shown that kNN-CDFs are a powerful way of analyzing large scale structure data sets. In particular, kNN-CDFs can break parameter degeneracies that are found with other statistical probes \cite{Banerjee_2022}. We investigate whether the response of kNN-CDFs to PNG is distinct from other processes and therefore whether kNN-CDFs can separate PNG from late-time non-Gaussianities generated by nonlinear gravitational evolution and galaxy formation. kNN-CDFs are closely related to the counts-in-cell (CiC) summary statistic. CiCs have been extensively studied \citep{Bernardeau_2000,Valageas_2002,Bernardeau_2015,Bernardeau_2016,Uhlemann_2016}, including for constraining PNG from the dark matter field \citet{Uhlemann_2018,Friedrich_2020}, and accurate analytical models have been developed for them \citep[see e.g.,][and references therein]{Uhlemann_2020}. By exploiting the relationship between CiCs and kNN-CDFs, we can obtain analytical models that describe our results and thereby replicate one desirable feature of bispectra analyses.

To examine the impact of primordial non-Gaussianity on kNNs we first use the \textsc{quijote-png} suite of simulations \citep{Coulton_2022a}. These simulations were designed to test PNG analysis methods and have been used to studying bispectrum statistics of the matter \citep{Coulton_2022a,Jung_2022a} and halo fields \citep{Coulton_2022b,Jung_2022b}, the halo mass function \citep{Jung_2023} and machine learning statistics \citep{Jung_2023,floss_2023}. Combined with the original \textsc{quijote} suite of simulations \citep{Villaescusa-Navarro_2020}, we can explore how kNNs respond to cosmological parameters jointly with  primordial non-Gaussianity.  

This paper is structured as follows: in \cref{sec:kNN_intro} we briefly review kNN-CDFs and in \cref{sec:simulations} we describe the simulations used in this work. In \cref{sec:halos} we apply kNN-CDFs to catalogs of dark matter halos and characterize the key features induced by PNG and their similarity to features arising from other key parameters. In \cref{sec:galaxies} repeat this analysis on a set of mock galaxy catalogs, compare the simulated catalogs to the CiC model and perform a Fisher forecast of the constraining power. We present our conclusions and outlook in \cref{sec:discussion}. In \cref{app:fixedDensity} we discuss how different choices in the definition of our sample impact the results and in \cref{app:convergence} we discuss the convergence of our numerical Fisher forecasts.

\begin{figure}
  \centering
  \includegraphics[width=.47\textwidth]{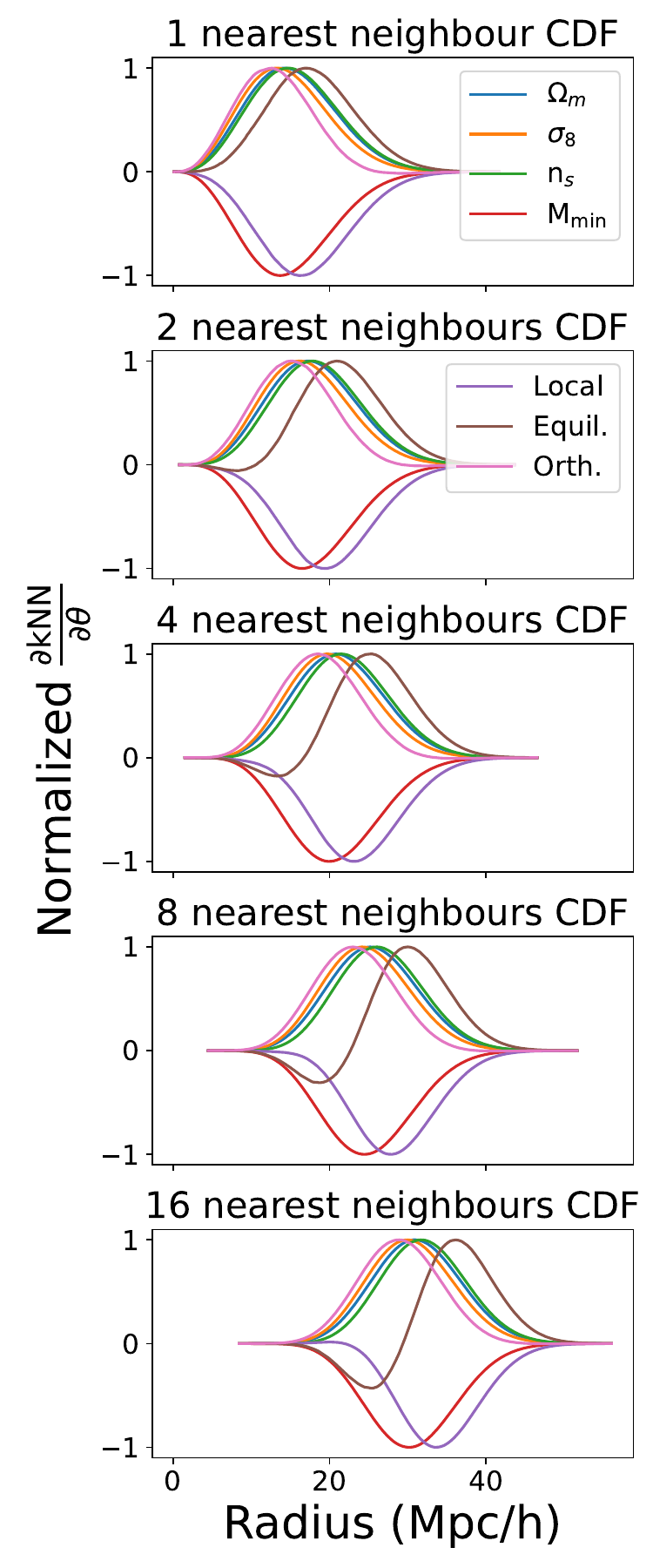} %height=.48\textwidth
\caption{ The normalized response of five dark matter halo kNN-CDFs to primordial non-Gaussianity,  cosmological and bias parameters. These responses are measured from the \textsc{quijote-png} simulations for all halos with $M_h\geq 3.2\times10^{13}$ M$_\odot$/h at $z=0$ and, for ease of comparison, each is normalized by its largest value. The response of the kNN-CDFs to \textit{equilateral} non-Gaussianity is different from the other parameters. }
\label{fig:param_effects_halos}
\end{figure}

\begin{figure*}
  \centering
  \includegraphics[width=.90\textwidth]{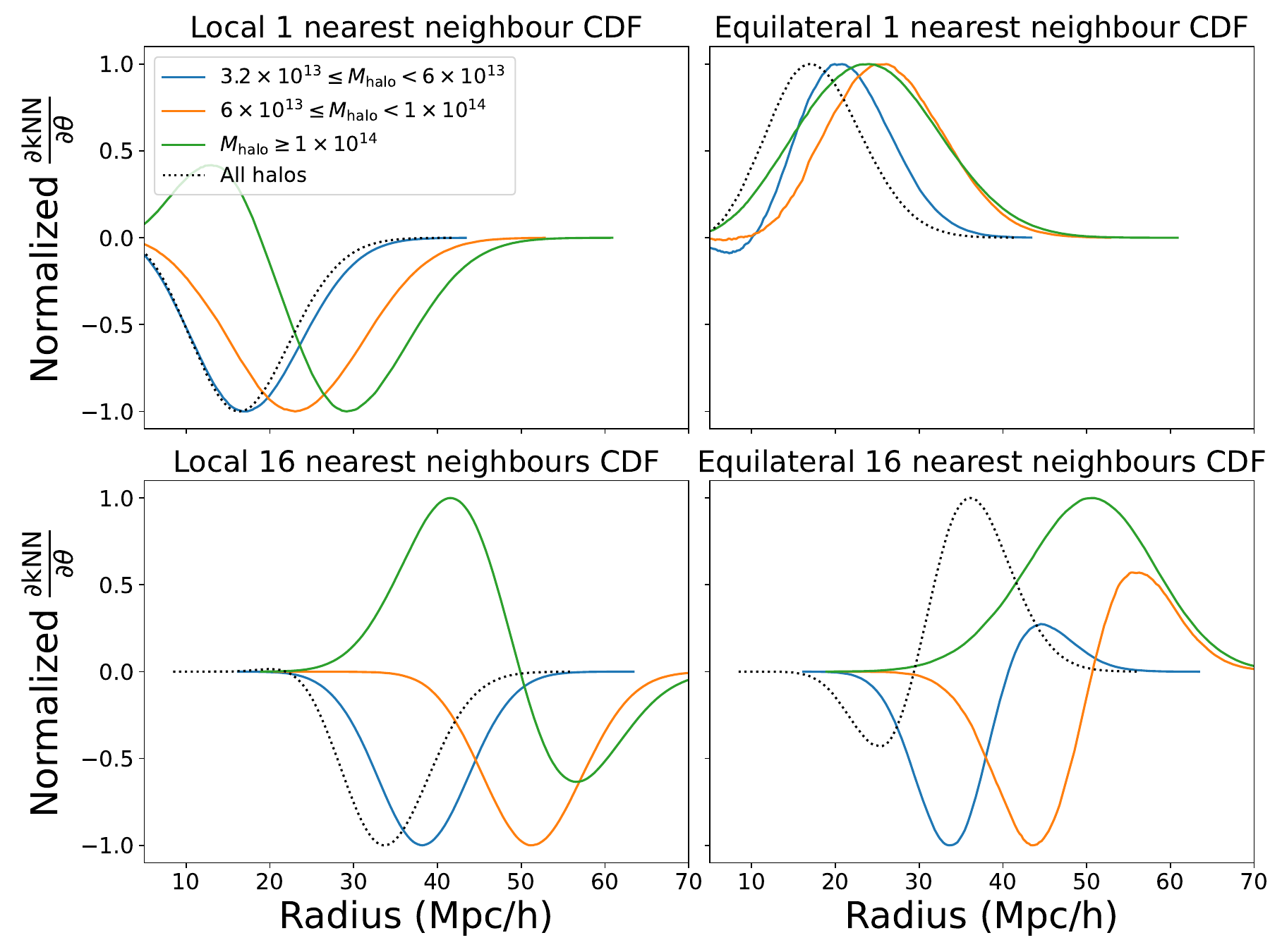} %height=.48\textwidth
\caption{ The signature of primordial non-Gaussianity in the kNN-CDFs displays a strong mass dependence as is demonstrated by examining three halo mass samples: a high mass sample ($M_h\geq 1\times 10^{14}$ M$_\odot$/h, green  ), an intermediate mass sample ($6\times10^{13}$M$_\odot$/h $\leq M_h< 1\times 10^{14}$ M$_\odot$/h, orange) and a lower mass sample  ($3.2\times10^{13}$M$_\odot$/h $\leq M_h< 6\times 10^{13}$ M$_\odot$/h, blue). For comparison we plot the total halo sample as a black dotted line.
\label{fig:halo_mass_dep}
}
\end{figure*}

\begin{figure*}
  \centering
  \includegraphics[width=.90\textwidth]{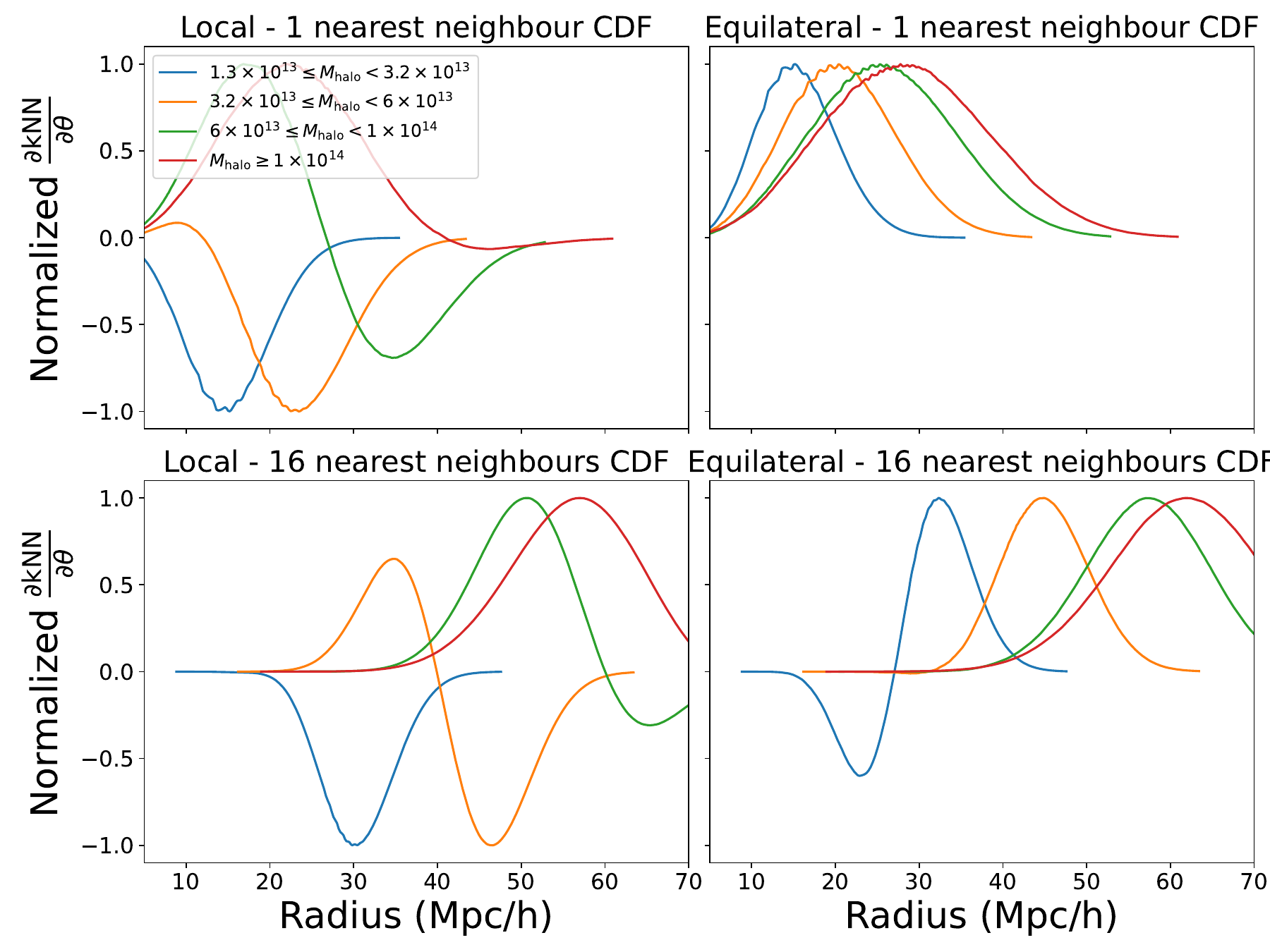} %height=.48\textwidth
\caption{ An examination of the kNN-CDFs obtained from halos at $z=0.5$ for four different halo mass samples. The configuration is otherwise the same as \cref{fig:halo_mass_dep}.
}
\label{fig:halo_mass_dep_z0p5}
\end{figure*}

\begin{figure}
  \centering
  \includegraphics[width=.47\textwidth]{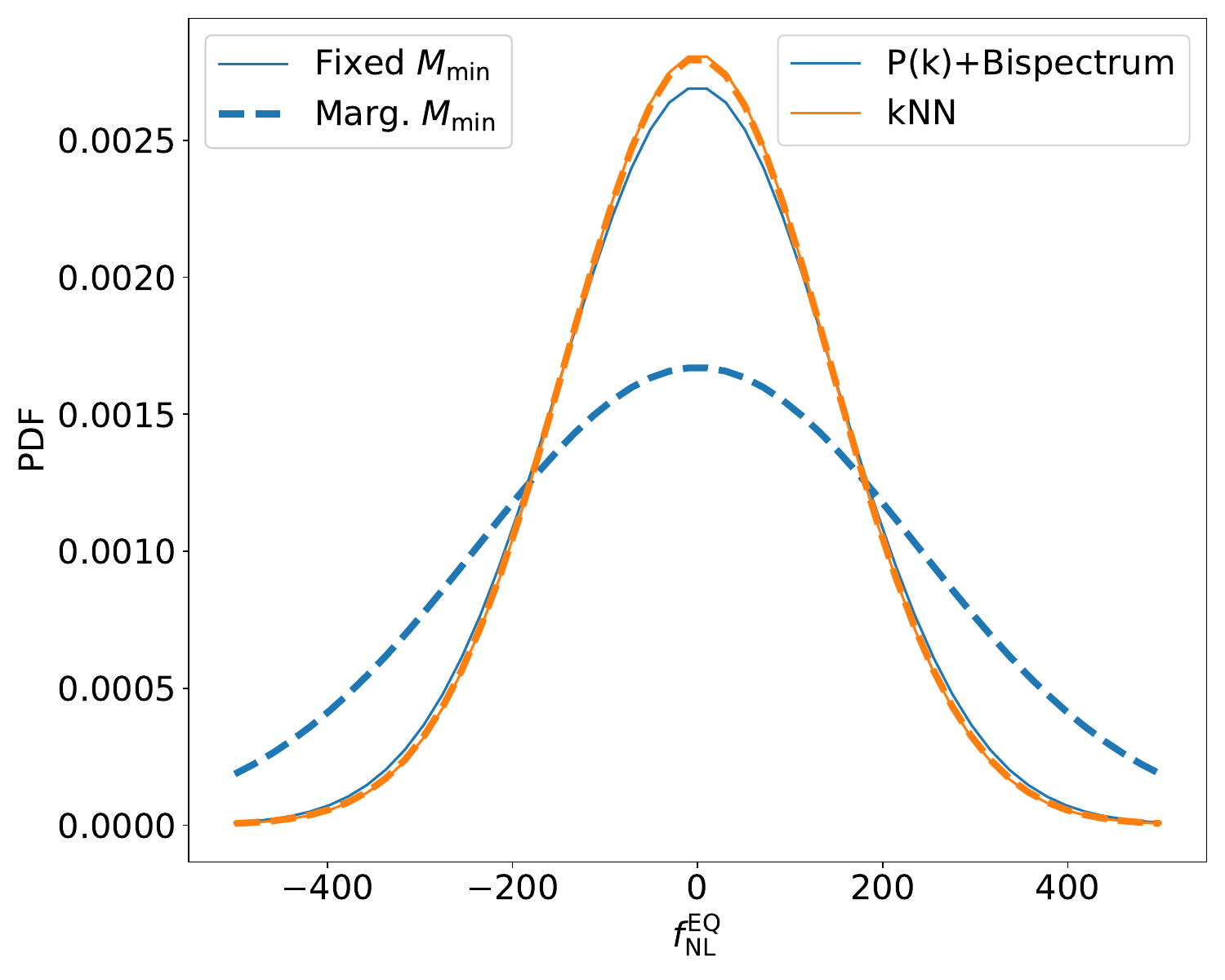} %height=.48\textwidth
\caption{ A comparison of the constraints on the amplitude of \textit{equilateral} non-Gaussianity, $f_\mathrm{NL}^\mathrm{equil.}$, from the bispectrum and power spectrum up to $k=0.5$h/Mpc \citep[as reported in][]{Coulton_2022b}, with those from the kNN-CDFs. This analysis uses the $M_h>3.2\times 10^{13}$ halo sample from the 1 Gpc$^3$ box at $z=0.0$ and considers two cases: constraining only $f_\mathrm{NL}^\mathrm{equil.}$ and constraining $f_\mathrm{NL}^\mathrm{equil.}$ whilst marginalizing an effective bias parameter, the minimum halo mass of the catalog $M_\mathrm{min}$. The distinct impact of primordial non-Gaussianity on the kNN-CDFs for this sample means that, unlike the bispectrum and power spectrum, the signature of primordial non-Gaussianity is not strongly degenerate with the bias parameter.
}
\label{fig:constraint_halo_impact_bias}
\end{figure}

\section{Overview of k-Nearest Neighbour cumulative distributions functions}\label{sec:kNN_intro}
k-Nearest Neighbour cumulative distributions functions simply measure the volume-averaged probability of finding at least $k$ objects with a sphere of radius, $R$. In this work we denote these statistics as ${\rm kNN}(R)$. They provide an alternative means of accessing the information contained within all the orientation averaged $N$-point correlation functions. There are several useful features of kNN-CDFs: first they can be computed in a very efficient manner \citep[see e.g.,][for details]{Banerjee_2021}. Second, they naturally can be applied to catalogs of objects, as is obtained from observations, without needing the data to be grided \citep[see e.g,][for issues arising from gridding data sets]{Jing_2005,Sefusatti_2016}. Third, the kNN-CDFs sample all regions of the volume equally rather than focusing on over-dense regions, yielding sensitivity to underdense regions in the volume. In fact, the $1{\rm NN}$-CDF is directly related to the Void Probability Function (VPF) \citep{White1979}.

The analysis here largely follows the methods described in \citet{Banerjee_2021} and we refer the reader to \citet{Banerjee_2021} for more details. To measure the CDFs we use the following procedure:
\begin{enumerate}
    \item We generate a set of distributed volume-filling points distributed on a grid. 
    We call these the query points. To ensure dense sampling, we typically use $10$ times as many query points as there are data points.
    \item We build a $k$d-tree using the data points in a chosen simulation (halo positions or galaxy positions). For each query point, we use this tree structure to compute the distances from the query points to the $k$-th nearest neighbor data points. In this terminology $k=1$ refers the closest data point to a query point, $k=2$ refers to the second nearest data point from a query point, and so on.
    \item For a particular $k$, we sort the list of distances to generate the empirical $k$NN-CDF for that $k$. We repeat the same step for different values of $k$.
    \item We then repeat the measurements for each simulation, and compute the average at a given cosmology and set of galaxy parameters.
\end{enumerate}
 To compute the response of the kNN-CDFs to different parameters we use finite differences as
\begin{align}
\frac{\partial \overline{kNN}}{\partial \theta} = \left<\frac{{kNN}|_{\theta=\theta+\delta\theta} -{kNN}|_{\theta=\theta-\delta\theta} }{2\delta\theta}\right>,
\end{align}
where $\delta\theta$ is a small step in the parameter of interest and the expected is computed as the average over the different simulation realizations.

The most significant difference between this work and  \citet{Banerjee_2021} is that we consider all objects in the catalog, rather using a fixed number of objects. This was used to allow easier comparison to previous \textsc{quijote-png} analyses \citep[e.g.,][]{Coulton_2022b}. In \cref{app:fixedDensity}, we present the results for samples with fixed number density and find a qualitatively similar picture.

\section{Simulations}\label{sec:simulations}
We use the \textsc{quiojte} and \textsc{quijote-png} simulations for a detailed description we refer the reader to \citet{Villaescusa-Navarro_2020,Coulton_2022a}. These are a suite of \textsc{gadget-3} \citep{Gadget} dark-matter, N-body simulations run at a cosmology consistent with \textit{Planck} \citep{planck2016-l06}: $\Omega_m=$0.3175, $\Omega_\Lambda=0.6825$, $\Omega_b=0.049$, $\sigma_8=0.834$, $h=0.6711$ and $n_s=0.9624$. The initial conditions were modified to include three shapes of primordial bispectrum:  \textit{local}, \textit{equilateral} and \textit{orthogonal}. The primordial bispectrum, $B_\Phi(\mathbf{k}_1,\mathbf{k}_2,\mathbf{k}_3)$, is defined as
\begin{align}
\langle \Phi(\mathbf{k}_1) \Phi(\mathbf{k}_2) \Phi(\mathbf{k}_3)  \rangle= \delta^{(3)}(\mathbf{k}_1+\mathbf{k}_2+\mathbf{k}_3) B_\Phi(\mathbf{k}_1,\mathbf{k}_2,\mathbf{k}_3)
\end{align}
where $\Phi(\mathbf{k})$ is the primordial potential at wavenumber $\mathbf{k}$. The three primordial bispectra considered here probe a range of different physical process \citep[see e.g,][for overviews]{Chen_2010,Meerburg_2019,Achucarro_2022} and have the forms:
\begin{align}
    B^{\mathrm{local}}_{\Phi}(k_1,k_2,k_3) = & 2 f_{\mathrm{NL}}^{\mathrm{local}} P_\Phi(k_1)P_\Phi(k_2)+  \text{ 2 perm.},
\end{align}
\begin{align}
  &   B^{\mathrm{equil.}}_{\Phi}(k_1,k_2,k_3) = 6 f_{\mathrm{NL}}^{\mathrm{equil.}}\Big[- P_\Phi(k_1)P_\Phi(k_2)+\text{ 2 perm.} \nonumber \\ &  
  -2 \left( P_\Phi(k_1)P_\Phi(k_2)P_\Phi(k_3) \right)^{\frac{2}{3}} +  P_\Phi(k_1)^{\frac{1}{3}}P_\Phi(k_2)^{\frac{2}{3}}P_\Phi(k_3)  \nonumber \\ & + \text{5 perm.}\Big],
\end{align}
and
\begin{align} \label{eq:bis_or_lss}
   & B^{\mathrm{ortho-LSS}}_\Phi(k_1,k_2,k_3) = \nonumber \\ & 6 f_{\mathrm{NL}}^{\mathrm{ortho-LSS}}
        \left(P_\Phi(k_1)P_\Phi(k_2)P_\Phi(k_3)\right)^{\frac{2}{3}}\Bigg[ \nonumber \\ &  -\left(1+\frac{9p}{27}\right) \frac{k_3^2}{k_1k_2} + \textrm{2 perms} +\left(1+\frac{15p}{27}\right)  \frac{k_1}{k_3} \nonumber \\ &   + \textrm{5 perms}  -\left(2+\frac{60p}{27}\right)  \nonumber \\ & +\frac{p}{27}\frac{k_1^4}{k_2^2k_3^2} + \textrm{2 perms}  -\frac{20p}{27}\frac{k_1k_2}{k_3^2}+ \textrm{2 perms} \nonumber \\ & -\frac{6p}{27}\frac{k_1^3}{k_2k_3^2} + \textrm{5 perms}+\frac{15p}{27}\frac{k_1^2}{k_3^2} + \textrm{5 perms}
    \Bigg],
\end{align}
where $P_\Phi(k)$ is the primordial potential power spectrum, $f_\mathrm{NL}^\mathrm{X}$ is the amplitude of each bispectrum and
\begin{align}
    p=\frac{27}{-21+\frac{743}{7(20\pi^2-193)}} \, .
\end{align}
We refer the reader to \citet{Coulton_2022a} for a detailed description of the implementation of these bispectra.

 For each shape 500 simulations are run with an amplitudes of the primordial bispectrum, $f_\mathrm{NL}^{X}= 100$, where $X$ denotes the shape, and 500 with $f_\mathrm{NL}^{X}= -100$. The seeds of the  $f_\mathrm{NL}^{X}= 100$ and  $f_\mathrm{NL}^{X}= -100$ simulations are matched to reduce cosmic variance. The \textsc{quiojte} simulations varied a set of cosmological parameters above and below the fiducial value, for use in Fisher forecasts. We used simulations that varied the amplitude of the linear matter fluctuations on smoothed on $8 {\rm Mpc}/h$ scales, $\sigma_8$, the Hubble constant, $h$, the fractional density of matter, $\Omega_m$, and the primordial spectral tilt, $n_s$. For each parameter there are 500 simulations with the parameter perturbed above and 500 perturbed below the fiducial value, again with matched seeds. We also use the 15,000 simulations run at the fiducial cosmology to compute covariance matrices.

For the analysis of the dark matter halos, we analyze the same samples as used \citet{Coulton_2022b,Jung_2022b}. Specially we use the friends-of-friends \citep[FoF,][]{FoF} halo catalog at redshifts, z, $z=0.0$ and $z=0.5$, and only include halos with mass, $M_h$, $M_h\ge 3.2 \times 10^{13}$ M$_\odot$/h. We work in redshift space by displacing the halos along the line of sight ($\hat{z}$ axis) according to their velocity. 

We use a halo occupation distribution (HOD) to generate mock galaxy catalogs from the simulations. Within the HOD framework used here, galaxies are assigned to halos probabilistically based solely on the halo mass, i.e. $P(N_\mathrm{gal}|M_h)$. In this work we use the \citet{Zheng_2007} formulation that decomposes the total number of galaxies in a halo into central and satellite contributions as
$N_\mathrm{gal} = N_\mathrm{central}+N_\mathrm{satellite}$. The central galaxies follow a Bernoulli distribution with mean
\begin{align}
    \langle N_\mathrm{central}\rangle = \frac{1}{2}\left[1+ \mathrm{erf}\left(\frac{\log M_h-\log M_\mathrm{min}}{\sigma_{\log M}} \right) \right]
\end{align}
and the satellite galaxies are Poissonian distribution with rate
\begin{align}
    \langle N_\mathrm{satellite} \rangle = \langle N_\mathrm{central}\rangle \left(\frac{M_h-M_0}{M_1} \right)^\alpha.
\end{align}
The parameters $M_\mathrm{min}$ and $\sigma_{\log M} $ set the minimum mass of halos that host galaxies and the width of the transition to hosting a central galaxy.  The parameters $M_0$, $M_1$ and $\alpha$ control the power law distribution of the satellite galaxies. The central galaxies are placed at the center of the halo, with the halo's velocity, whilst the satellite galaxies are distributed according to a NFW profile with velocities set acording to the isotropic Jeans equations \citep{Navarro_1996,lokas_2001}. We use these velocities to displace the galaxies along the line of sight to produce catalogs in redshift space.

\citet{Biagetti_2023} derived a set of best fit HOD parameters for the \textsc{quijote-png} simulations, such that the galaxy catalogs matched the  CMASS BOSS galaxy survey at $z=0.5$. 
We used a set of parameters motivated by those fits: $M_\mathrm{min}=2.2\times10^{13}$ M$_\odot$/h, $ M_0 = 2.8\times 10^{13}$  M$_\odot$/h, $ M_1 = 1.78\times 10^{14}$  M$_\odot$/h, $\sigma_{\log M}=0.15$ and $\alpha=0.5$. We use these parameters to generate catalogs at $z=0.0$. The purpose of this choice is to demonstrate properties of the kNN-CDFs and provide a comparison to previous works at  $z=0.0$. Thus this galaxy catalog is not designed to match any specific experiment. Note that minimum dark matter halo mass used for the galaxy catalogs is $M_h=1.3\times 10^13 M_\odot/h$, so lower than that used in analyses of the dark matter halos. 

\section{The impact of PNG on dark matter halo k-Nearest Neighbour cumulative distribution functions}\label{sec:halos}
In \cref{fig:param_effects_halos} we show how kNN-CDFs distributions for five different numbers of neighbours respond to PNG, variations in cosmological parameters and a simple bias parameter ($M_\mathrm{min}$). Interestingly the kNN-CDFs statistics, when compared across different numbers of nearest neighbours, respond differently to \textit{equilateral} non-Gaussianity than to all other parameters.

To understand this further we break the halo catalog into three subsets: a high mass sample, $M_h\geq 1\times 10^{14}$ M$_\odot$/h, an intermediate sample, $6\times10^{13}$M$_\odot$/h $\leq M_h\leq 1\times 10^{14}$ M$_\odot$/h , and a low mass sample $3.2\times10^{13}$M$_\odot$/h $\leq M_h\leq 6\times 10^{13}$ M$_\odot$/h. The results are shown in  \cref{fig:halo_mass_dep}. There is a complex mass dependence of these responses and for some mass bins the \textit{local} PNG shows a similar response. These results show similarities to effects seen in the halo mass function \citep[see e.g.,][]{LoVerde_2008,Wagner_2010,Jung_2023}; the response of the halo mass function to \textit{equilateral} and \textit{local} PNG changes sign at $~7\times10^{13} M_\odot$/h and  $1\times10^{14} M_\odot$/h. These similarities suggest a common underlying cause. Changes in the number density will impact the kNN-CDFs; the simplest case of a rescaling of the mass function leads to a horizontal shift of the kNN-CDF. The response of the halo mass function to PNG is complex \citep[see e.g, Fig 1 of ][]{Jung_2023} with the number of high mass halos being boosted, whilst the number of low mass halos is reduced.  As different mass halos are clustered to differing extents we expect a complex signature in the kNN-CDFs. A second suggestive piece of evidence to support this hypothesis is that the redshift evolution of the kNN signature mirrors the effects seen in the mass function. This can be seen in \cref{fig:halo_mass_dep_z0p5} where we show the response of the kNN-CDFs at redshift $z=0.5$, again split in four mass samples. The distinctive feature moves to lower mass, as does the signature in the halo mass function.

A key challenge in constraining \textit{equilateral} non-Gaussianity is disentangling it from non-Gaussianity introduced by the non-linear evolution of the LSS. The distinct impact of PNG on the kNN-CDFs, for certain samples, means that the degeneracy with these late time effects will be significantly reduced. To explore this we perform a simple Fisher forecast for constraints using kNN-CDFs measurements. In this forecast we use kNN-CDFs with the following number of neighbours: 1, 2, 4, 8, 16, 32, 64 and 128. We cut the kNN-CDFs at a minimum scale of 10 Mpc/h and cut the tails of the distributions, where kNN-CDF$<0.005$ or kNN-CDF$>0.995$. We use the halo catalog with $M_h>3.2\times10^{13}$M$_\odot$/h at z=0.0.

An interesting question is what likelihood describes the kNN-CDFs. Characterizing the distribution of the kNN-CDFs is complex and in this work we consider an alternative avenue. We choose to compress the statistics and then assume a Gaussian likelihood for the compressed statistics. From other studies \citep{Anbajagane2023} , we know that the likelihoods of the CDFs are very close to Gaussian, as long as we stay away from the tails and do not sample the CDF too densely. If the likelihood of the kNN-CDFs was known, the data could be compressed lossesly into a set of summary statistics that, as quasi maximum likelihood estimators, are Gaussian distributed \citep[see e.g.,][]{Lehmann_2006,Alsing_2018}. Here compress the kNN-CDFs distribution functions using the \textsc{moped} compression \citep{Heavens_2000} as
\begin{align}
\hat{\theta}_i = \frac{\partial \overline{kNN}} {\partial \theta_i}\mathcal{C}^{-1} \left(kNN - \overline{kNN}\right),
\end{align}
where $\hat{\theta}_i$ are the compressed statistics and $\overline{kNN}$ and $\mathcal{C}$ are the mean and covariance of the kNN-CDFs measurements. As the kNN-CDFs are not Gaussian, this compression losses information. However, the compressed statistics are well approximated by a Gaussian distribution - this can be understood through the central limit theorem.
We then compute forecast parameter constraints as
\begin{align}
    \sigma(\theta_i)^2 = F^{-1}_{ii}
\end{align}
where the Fisher information is given by
\begin{align}
    F_{ij} = \frac{\partial \hat{\theta_I}}{\partial{\theta}_i}\Sigma_{IJ}\frac{\partial \hat{\theta_J}}{\partial{\theta}_j},
\end{align}
and $\Sigma_{IJ}$ is the covariance of the summary statistics. We compute the derivatives for compression and the Fisher forecast numerically as described in \citet{Coulton_2022a}. We split the simulations into two disjoint sets: the first set is used for the compression and the second for the Fisher forecast. A more detailed discussion of Fisher forecasts using compressed statistics is given in \citet{Coulton_2023b}.

As seen in \cref{fig:constraint_halo_impact_bias}, marginalizing over a simple bias parameter dramatically degrades the power-spectrum and bispectrum prediction, whilst leaves the kNN-CDF result largely unchanged. It is well known that the impact \textit{equilateral} non-Gaussianity of the galaxy power-spectrum and bispectrum is highly degenerate with galaxy bias \citep{Baldauf_2016,Lazanu_2017,Baumann_2022}; our results, even with a simple bias model, reproduce this. This degeneracy significantly degrades our constraints and limits the scales that can be used. Effective field theory (EFT) approaches provide a systematic way to marginalize over these effects \citep{Baumann_2012}, however no systematic method exists for scales beyond the validity of EFT and the large degeneracy means obtaining a robust result from non-linear scales will be difficult. On the other hand, this result demonstrates that kNN-CDFs can effectively disentangle this bias parameter and PNG. This suggests that they many provide a path to robust, small-scale PNG measurements and in the next section we test this hypothesis more stringently with an extended, and more realistic, bias model. It is also interesting that the unmarginalization bispectrum and kNN constraints are very similar.

\section{Extracting signatures of PNG from galaxy k-Nearest Neighbour distributions}
\label{sec:galaxies}
Next we apply the kNN-CDFs to the mock galaxy sample described in \cref{sec:simulations}. The choice of this galaxy sample is motivated by the results of \cref{sec:halos}: it is interesting to see whether, for a galaxy sample not dominated by contributions from halos with $M_h>10^{14}$ M$_\odot$/h, the \textit{equilateral} PNG signature remains.

In this section, we primarily focus on exploring what can be learnt about \textit{equilateral} non-Gaussianity with kNN-CDFs. 

\begin{figure*}
  \centering
  \includegraphics[width=.99\textwidth]{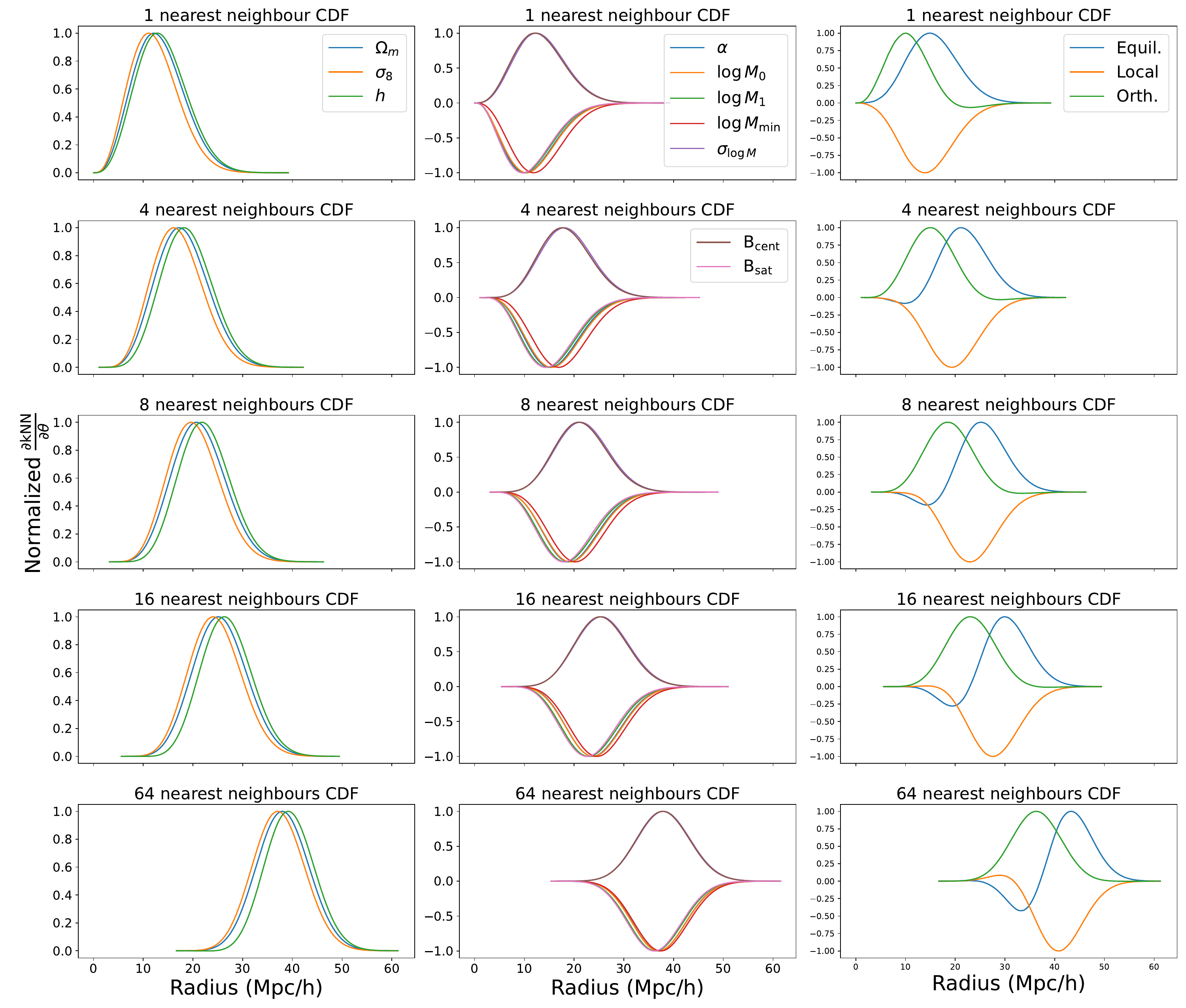} 
\caption{ The signature of primordial non-Gaussianity, cosmological and HOD parameters in the kNN-CDFs of the galaxy sample. The response of this galaxy sample to primordial non-Gaussianity is distinct from the response to cosmological and HOD parameters.
}
\label{fig:param_galaxies_effects}
\end{figure*}

\subsection{The impact of PNG on galaxy kNN-CDFs}

Using the HOD galaxy sample we compare how cosmological parameters, HOD parameters, and primordial non-Gaussianity impact the properties of the galaxy sample. As seen in \cref{fig:param_galaxies_effects}, the impact of \textit{equilateral} PNG in this sample remains distinct from all other contributions. Note that several of the HOD parameters are highly degenerate. In particular $\sigma_{\log M}$ and $\log M_\mathrm{min}$ are essentially perfectly degenerate.

The fiducial HOD used in this work, as described in \cref{sec:simulations}, assumes that the number of galaxies in each dark matter halo is only a function of the halo's mass. However, numerous studies of hydrodynamical simulations and semi-analytic models have shown that other properties, such as concentration and environment, are important \citep[e.g.,][]{Gao_2005,Gao_2007,Wechsler_2006,Croton_2007,Li_2008,Bose_2019,Hadzhiyska_2020,Xu_2021,Hadzhiyska_2021}. The dependence on additional halo properties is called `assembly bias'. Observational evidence that assembly bias is important for current two-point analyses is mixed: some works find assembly bias to be important \citep{Zentner_2019,Yuan_2021,Yuan_2022,Contreras_2023} and others finding no strong evidence \citep{Lin_2016,Niemiec_2018,Salcedo_2022,Yuan_2023b,Rocher_2023}. The reconciliation of these mixed results lies in the scales used in the analyses and the properties of the samples. Given the increasing sensitivity of upcoming surveys, the importance of assembly bias in simulations and small scale observations, and the hints that assembly bias may be more important for beyond two-point statistics \citep{Yuan_2023}; we explore how one model of assembly bias impacts our kNN-CDF analysis.

We explore assembly bias using the Luminous Red Galaxy (LRG) extended HOD from \citet{Yuan_2022,Yuan_2023}. This model uses the local density around dark matter halos as the secondary halo parameter that modulates the number of halos within a dark matter halo. This is implemented by the following modifications
\begin{align}
   & \log M^{\mathrm{new}}_\mathrm{min} = \log M_\mathrm{min}+B_\mathrm{cent}\left(\delta_m^\mathrm{rank}-0.5\right), \nonumber \\
   & M^\mathrm{new}_0 = \frac{M_0}{M_\mathrm{min}}M^\mathrm{new}_\mathrm{min}, \text{\, and} \nonumber \\
   &\log M_1^\mathrm{new} =\log M_1 +B_\mathrm{sat}\left(\delta_m^\mathrm{rank}-0.5\right),
\end{align}
where $B_\mathrm{cent}$ and $B_\mathrm{sat}$ characterize the level of assembly bias and $\delta_m^\mathrm{rank}$ is the rank of the local over-density, smoothed on a scale of $5$Mpc/h.

In \cref{fig:param_galaxies_effects}, we show how response of the kNN-CDF measurements to variations of the two assembly bias parameters, around the fiducial values of 0. We see that, for this galaxy sample, the impact on the kNN-CDFs measurements is not similar to the signature of \textit{equilateral} non-Gaussianity. However, further work is required to test whether this holds for other galaxy samples and other secondary parameters (such as concentration or local tidal field). 

\subsection{Analytical Modeling}\label{sec:cic_model}
kNN-CDFs are closely related to counts-in-cell statistics, which characterize the distribution of number counts, or the over density, smoothed over some scale. More precisely, measuring all the kNN-CDFs for all numbers of nearest neighbours at a single radius is exactly equivalent to the pdf of counts-in-cell (CiC) statistic, smoothed with a spherical top hat, evaluated a single smoothing scale.  Thus, a measurement of all nearest neighbours of the kNN-CDFs is exactly equivalent to measuring the CiC pdf at all smoothing scales. However, typical kNN-CDFs analyses only consider a subset of all the kNN-CDFs nearest neighbours distributions (here we have used nearest neighbours 1-8,16,32,64 and 128), whilst typical CiC analyses consider only a handful of smoothing scales. From this view, the two statistics are highly complementary.

There has been extensive work on CiC statistics \citep[e.g.,][]{Fosalba_1998,Valageas_2001,Valageas_2002a,Bernardeau_2014,Uhlemann_2016,Ivanov_2019,Uhlemann_2020} including on using CiCs measurements for constraining PNG \citep[e.g,][]{Gaztanaga_1998,Valageas_2002b,Uhlemann_2018,Friedrich_2020}. Most recently \citet{Friedrich_2020} found that the counts-in-cell matter field pdf is a powerful probe of primordial non-Gaussianity that can be modelled analytically. This model for the matter field can be combined with work on the galaxy-matter connection to obtain predictions for the galaxy CiC pdf. In this work we use the \textsc{cosmomentum} package developed in \citet{Friedrich_2020} with the galaxy-matter connections from \citet{Friedrich_2018,Friedrich_2022}.

These CiC analytical advancements can be equally applied to understand kNN-CDFs measurements, thanks to the intimate connection between the two statistics. As analytical modeling of the kNN-CDFs has been considered in past work \citep{Banerjee_2022}, here we simply use the CiC models to validate our results. To do so we map the kNN-CDF measurements onto the CiC framework and compare our measurements to the CiC predictions. We perform this comparison, rather than mapping the analytical predictions on the kNN-CDF frame, primarily as the model requires a rescaling of the non-linear variance at each smoothing scale, see \citet{Friedrich_2020} for a detailed discussion of this. This rescaling is performed once for the CiC pdf, as it is a single smoothing scale, but must be performed for each radius of the kNN-CDFs. As this equates to a rescaling of each point for a single kNN-CDFs, it is only through consistency across kNN-CDFs, with many different numbers of nearest neighbours, can the theory model be rigorously tested. Tests across different numbers of nearest neighbours of kNN-CDFs effectively equates to the CiC picture!  A more minor reason is that the analytical method is only valid for the bulk on the CiC pdf and these cuts have a simpler expression in the CiC framework.

To generate our analytical predictions we use the Lagrangian bias method from \citet{Friedrich_2022}, matching the galaxy linear bias, $b_1$, and number density, $\bar{n}$ to that of our simulated galaxy catalogs. Further we use our simulations to compute the non-linear variance for each smoothing scale and use this to calibrate our analytical predictions as described in \citet{Friedrich_2020}. We refit the parameters of the non-Poissonian shot noise \citep[c.f. Eq. 24 of][]{Friedrich_2022} to match our simulations. We find the following parameterization of the non-Poissonian shot noise describes the bulk-pdf behavior of our simulations
\begin{align}
\frac{\langle N^2|\delta_m \rangle-\langle N|\delta_m \rangle^2}{\langle N|\delta_m \rangle} = \begin{cases}
			\alpha_0+\alpha_1 \delta_m, & \text{if $\delta_m<0$}\\
            \alpha_0, & \text{otherwise},
		 \end{cases}
\end{align}
where $\alpha_0=0.95$ and $\alpha_1=-0.35$ control the deviation from Poissonian noise.
An alternative could be a quadratic bias model, however for simplicity we use piece-wise linear model. 

\begin{figure}
  \centering
  \includegraphics[width=.47\textwidth]{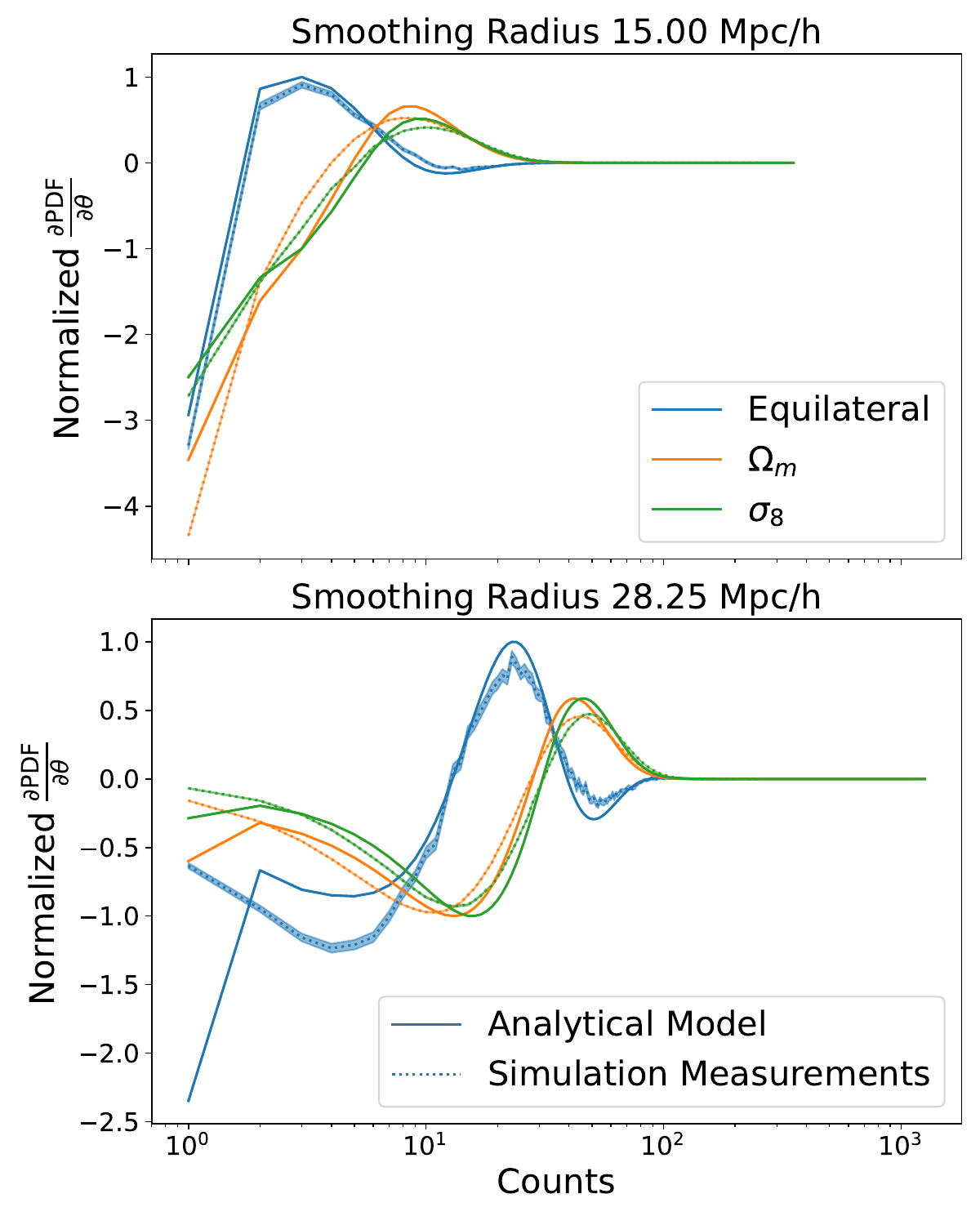} 
\caption{ A comparison between the analytical model presented in \cref{sec:cic_model}, solid lines, and the kNN-CDF measurements converted to counts-in-cells for two smoothing scales, dotted lines. The error bars denote the error on the mean.
}
\label{fig:cic_responses}
\end{figure}

In \cref{fig:cic_responses} we compare measurements of the derivatives of the first 250 kNN-CDFs of our galaxy sample to the analytical predictions.  For both the smoothing scales shown, $15$\,Mpc/h and $28.25$\,Mpc/h, and for all three parameters ($\sigma_8$, $\Omega_m$ and $f_\mathrm{NL}^\mathrm{equil.}$) we find generally good agreement between the simulations and theoretical model. 
This agreement provides strong evidence that the signatures of PNG in the kNN-CDFs are physical, and not an artifact of our simulation methods. This eases any potential concerns on the limitations of the simulations, such as on the impact of resolution, finite volumes, the approximation generation of primordial non-Gaussianity or missing components (such as baryons). The analytical method also helps develop intuition behind our observed features.

The theoretical model from \citet{Friedrich_2020} and \citet{Friedrich_2022} focused on modelling the bulk of the pdf and thus the model is not expected to work in the tails of the distribution. Further, \citet{Friedrich_2020} noted that resolution effects can lead to small discrepancies between the model and simulations. Combined these effects are thought to explain the observed differences seen in our measurements.

\subsection{Parameter Constraints}\label{sec:param_constraints}
\begin{figure*}
  \centering
  \includegraphics[width=.95\textwidth]{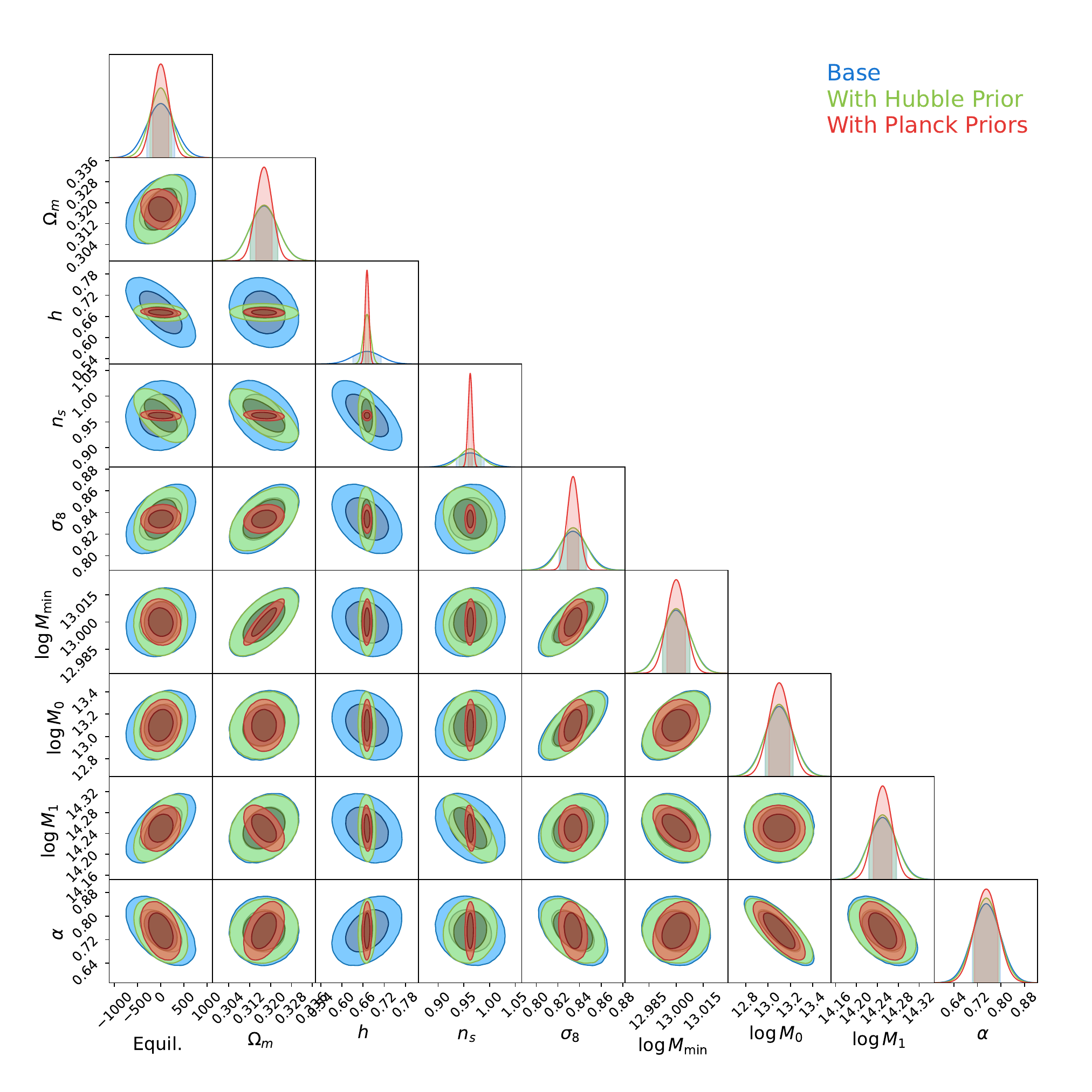} 
\caption{ Constraints on \textit{equilateral} non-Gaussianity and a set of cosmological and HOD parameters. Whilst the impact of \textit{equilateral} non-Gaussianity is not degenerate with any single parameter, there are more complex degeneracies in the large 9 dimensional parameter space. As the kNN-CDF information primarily comes from small-scale, $<50$ Mpc/h, clustering these measurements can easily be combined with probes of the background expansion, large-scale clustering and CMB measurements. Thus, we compare constraints from only the kNN-CDFs (blue) with constraints including a prior on the Hubble constant consistent with local distance measurements \citep{Riess_2022} or a \textit{Planck} 2018 prior on the cosmological parameters ($\sigma_8$, $\Omega_m$, $h$ and $n_s$). Combining just with the Hubble prior allows most of the \textit{equilateral} degeneracies to be removed. These constraints are from our galaxy sample at $z=0.0$ with a volume of $1$ Gpc$^3$.
}
\label{fig:param_constraints}
\end{figure*}

Using these results we consider a Fisher forecast using the galaxy sample. As in \cref{sec:halos} we perform the Fisher forecast in the compressed data space. As $\sigma_{\log M}$ and $\log M_\mathrm{min}$ are perfectly degenerate for our sample, we only consider one of these as a free parameter, $\log M_\mathrm{min}$ . The results are shown in \cref{fig:param_constraints}.  From 1Gpc$^3$ volume the kNN-CDFs would be able to constrain \textit{equilateral} non-Gaussianity to $\sigma(f_\mathrm{NL}^\mathrm{equil.})=299$. We note that the resolution limitations of the simulations mean that a realistic sample, which much higher number densities, would be able to obtain tighter constraints. The value of this approach can be seen by comparing to results in \citep{Coulton_2022b,Jung_2022b,Jung_2023}, where the sample halo sample is analyzed with bispectrum and halo mass function statistics.

Whilst $f_\mathrm{NL}^\mathrm{equil.}$ is not degenerate with any single parameter, see \cref{fig:param_galaxies_effects}, there are some mild degeneracies in the 9 dimensional space. Quantitatively, the marginalized constraint is 2.4 times larger than the case when the eight other parameters are fixed. Given the large degree of flexibility within the 9 parameter space, it is not  surprising that degeneracies appear. The degradation seen here is similar to that found when marginalizing over the same cosmological parameters and only one, simple bias parameter in a bispectrum halo analysis \citep{Coulton_2022b,Jung_2022b}, implying that the size of kNN degeneracies is significantly smaller. 

Given the large parameter space, it is unsurprising that parameter degeneracies would develop. The information in the kNN-CDFs primarily comes from small-scale clustering. Therefore, a kNN-CDF analysis could be combined with other cosmological probes such as baryon acoustic oscillation measurements, supernova, CMB power spectra measurements or even large-scale galaxy power spectrum and bispectrum measurements. These probes are all highly complementary and could be used to break degeneracies with the kNN-CDFs. In \cref{fig:param_constraints} we show two examples of how this could be beneficial. In the first case we add a prior on the Hubble parameter, this prior is from \citet{Riess_2022} and in the second case we include independent priors on the cosmological parameters based on \textit{Planck} 2018 cosmology \citep{PlanckVI_2018}. In both of these cases we see significantly reduced degeneracies $f_\mathrm{NL}^\mathrm{equil.}$. With the \textit{Planck} priors the $f_\mathrm{NL}^\mathrm{equil.}$ constraint is only $40\%$ larger than the case where all other parameters are fixed. This demonstrates that kNN-CDFs can separate late-time physics, such as how galaxies occupy halos, from the primordial signatures.  Note that these priors are purely demonstrative, future experiments would offer significantly improved priors. 

\section{Discussions}\label{sec:discussion}
In this work we have simulated the impact of three types of primordial non-Gaussianity on dark matter halo and galaxy kNN-CDFs from the \textsc{quiojte-png} simulation suite. The signatures of primordial non-Gaussianity in kNN-CDFs are distinct from changes in cosmological parameters and also from bias parameters (a simple $M_\mathrm{min}$ for dark matter halos and five halo occupation distribution parameters for the galaxies). The signature of PNG was characterized by examining multiple mass samples and two redshifts. The mass and redshift evolution show features similar to the halo mass function and therefore suggest the kNN-CDFs are accessing related physical effects.  The scales used in this analysis ($10-70$ Mpc/h) are smaller than typical primordial non-Gaussianity analyses \citep{Nishimichi_2020,DAmico_2022b,Philcox_2022} and so present a highly complementary analysis approach. An interesting future topic would be to combine these approaches. This would utilize both the optimal analysis of the large scales with the bispectrum and  small scale information probed by the kNN measurements. Further kNN-CDFs have already been shown to powerfully constrain bias parameters \citep{Yuan_2023} and this could further enhance large-scale bispectrum measurements.

To further validate and build intuition for our results, we compared them to theoretical predictions. This was done by exploiting the close relationship between kNN-CDFs and counts-in-cells. The impact of PNG on counts in cells for dark matter has been studied before \citep{Uhlemann_2018,Friedrich_2020} and accurate analytic tools have been developed for those results. Mapping our kNN-CDFs results into the counts-in-cells (CiC) frame we found reasonable good agreement with the theoretical prediction. This provides a stringent validation of simulations and helps demonstrate that spurious artifacts of the initial conditions, which can arise from higher order effects of the IC generation method, are negligible. Further, this provides a theoretical framework to further understand the signatures found in the simulations.  The close relationship between kNN-CDFs and CiCs means that our results can equally be thought of as both an extension of the results of \citet{Friedrich_2020} to dark matter halos and galaxies and an examination of the CiC space (counts at fixed scales) from a different perspective (thresholded counts as a function of scale).

In the primordial space, the primordial non-Gaussianities are bispectra as a function of both scale and configuration -- the relative magnitudes of the three wavevectors. The theoretical model that accurately describes our kNN-CDFs is only a function of the skewness at different scales of the linear density field, as demonstrated in Eq. 52 of \citet{Friedrich_2020}. This has two consequences for the kNN-CDFs: first, this states that distinct signature of \textit{equilateral} PNG, shown in \cref{fig:param_effects_halos} and \cref{fig:param_galaxies_effects}, arises due to the scale dependence of the \textit{equilateral} PNG skewness. This is connected to the halo mass function, which shows similar mass dependence and is discussed in \cref{sec:halos}, as the skewness on different scales controls the response of the halo mass function to PNG \citep{Chongchitnan_2010,LoVerde_2011}. Second, this suggests that kNN-CDFs are not able to fully access all the information contained in the primordial bispectrum, as the skewness averages over the configuration of the bispectrum and may ``wash out" some of the signal. An interesting question is how much of the primordial signal is lost. To answer that we compare the information in the skenwness of the linear density field to the bispectrum of the same field. The results, shown in \cref{tab:skewness_vs_bispec}, indicate that for \textit{equilateral} non-Gaussianity almost all of the information can be accessed via measurements of the skewness. For the other types of non-Gaussianity, especially \textit{orthogonal}, more information is lost. 

There are several interesting future directions. First ,the galaxy sample considered here does not represent an observational sample. It was constructed to explore the HOD degeneracies for a sample populating the `low mass' halos in our simulation. Our simulation mass resolution leads to a minimum halo mass is very close to the mean mass of observational samples \citep[e.g., BOSS, unWISE, DESI LRGs  and ][]{More_2015,Krolewski_2020,Yuan_2023b}. Considering a more realistic sample, which requires higher resolution simulations, would be a valuable next step. Second, whilst the catalogs used in this analysis were in redshift space, the kNN-CDFs used here were isotropic. Recent work by \citet{Yuan_2023} has shown that using 2 dimensional kNN-CDFs allows more information to be extracted from cosmological data sets. A third interesting extension, also proposed in \citet{Yuan_2023} , is to compute the kNN-CDFs not from a set of random points, but from the data-points themselves. This statistic was shown to be highly effective in constraining HOD parameters and thus could be effective for further reducing degeneracies between PNG and late-time physics. The results in the main text used all the halos/galaxies in the catalog. This induced a dependence on the total number of objects in the sample. In \cref{app:fixedDensity} we explore one means of obviated this: using a fixed number density. However, for a realistic survey we would desire to use all observed objects (rather than downsampling to a desired number density). An alternative would be to normalize the radius by the mean separation.  Similarly this analysis worked in the simplified geometry of a periodic box at a single redshift. Exploring the impacts of non-trivial geometries, light-cone effects \citep[e.g.][]{Yuan_2023c}, observational masks and combining multiple redshifts would set the stage for an analysis of observations. A final direction would be to consider in more detail the effects of assembly bias -- for example including velocity bias or investigating alternative secondary halo parameters. These topics will be the subject of future work.  

In conclusion, these results suggest that kNN-CDFs could be a powerful statistic for separating out primordial non-Gaussianity and late-time physical processes.

\section*{Acknowledgements}
The authors are very grateful to Sandy Yuan, Oliver Philcox, Francisco Villaescusa-Navarro, David Spergel, Oliver Friedrich, Cora Uhlemann and the \textsc{Quijote-PNG} collaboration for useful discussions. This work was supported by collaborative visits funded by the Cosmology and Astroparticle Student and Postdoc Exchange Network (CASPEN). This work was also supported by U.S. Department of Energy grant DE-AC02-76SF00515 to SLAC National Accelerator Laboratory managed by Stanford University. 
\begin{table}
    \centering
    \begin{tabular}{c c c}
& \multicolumn{2}{c}{1$\sigma$ Fisher constraint} \\
PNG Type & Bispectrum & Skewness \\
    \hline \hline 
\textit{Local} & 4.4 & 6.1 \\
\textit{Equilateral} & 18 & 20 \\
\textit{Orthogonal} &8.0 & 15 \\
   \end{tabular}
    \caption{A comparison of the constraining power obtainable from measurements of the skewness, as a function of scale, and the full bispectrum of the linear field at $z=0.0$. These results are obtained from an analytic Fisher forecast, with a $k_\mathrm{max}=0.5$h/Mpc, for both analyses.\label{tab:skewness_vs_bispec} }
\end{table}

\appendix
\section{Fixed Number Density}\label{app:fixedDensity}
\begin{figure}
  \centering
  \includegraphics[width=.47\textwidth]{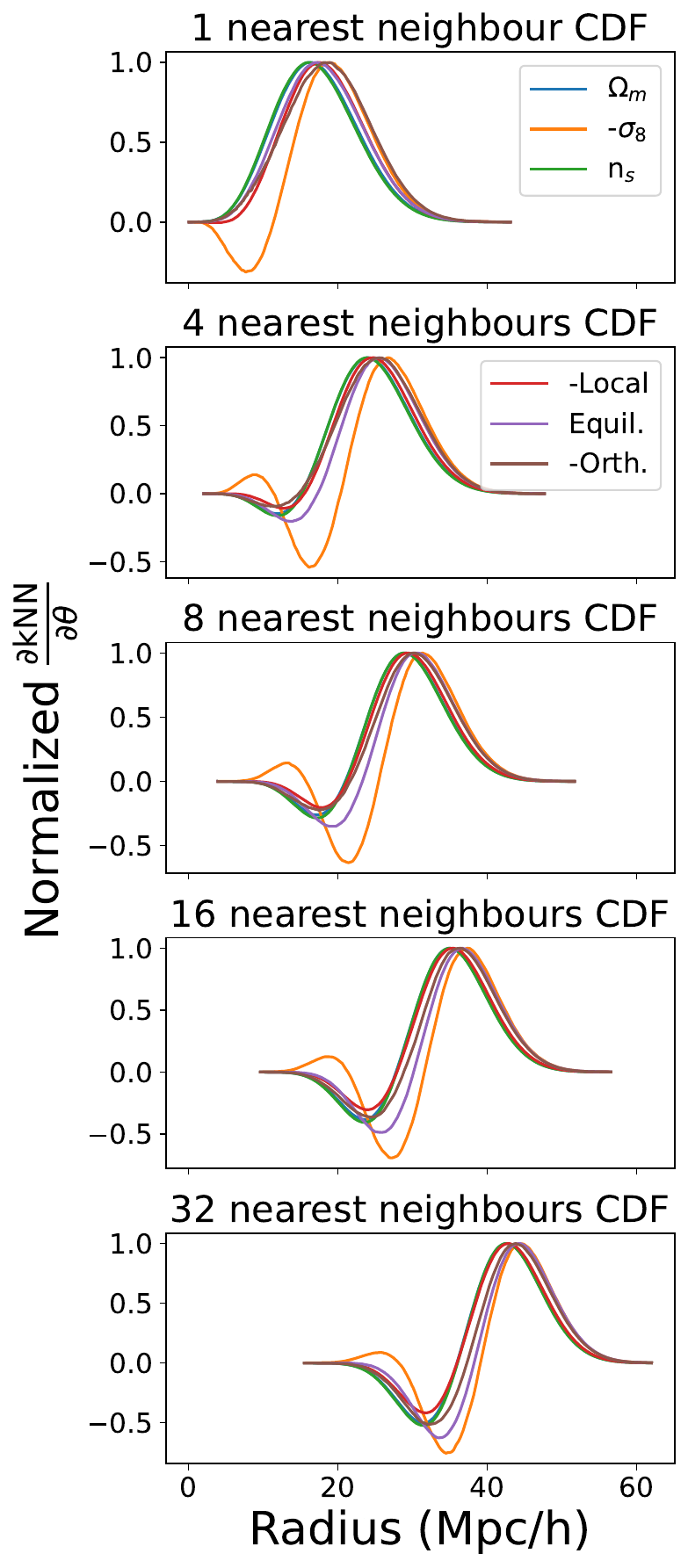} %height=.48\textwidth
\caption{ The kNN-CDFs of the halo field, as computed as in Fig. \ref{fig:param_effects_halos}, however these are evaluated at fixed number density of $n_h=0.00015$ (h/Mpc)$^3$. Note that to aid comparison of the derivatives they have all been normalized by the peak value. The signature of primordial non-Gaussianity is still distinct from the other parameters. This shows that effect does not arise solely from a change in the total number of halos. }
\label{fig:param_effects_halos_fixed}
\end{figure}

In this appendix we investigate how a different analysis choice, computing the kNN-CDFs from a fixed number of data points, alters our conclusions. Using a fixed number of objects, rather than all objects, removes the sensitivity to the total number of objects. In \cref{fig:param_effects_halos_fixed} we recompute the dark matter halo kNN-CDFs using samples of 150000 randomly chosen dark matter halos at $z=0.0$ with $M_h>3.2\times 10^{13}$\,$M_\odot/h$. The response of all kNN-CDFs to different parameters is different to the case where all the halos are used. The smooth bell curves are no longer seen as the shifts in the number of halos has been removed. Despite the difference in shape, the response to \textit{equilateral} non-Gaussianity across the different numbers of neighbours is still distinct in shape from the other cosmological parameters. In \cref{sec:halos}, we discussed that the kNN-CDFs may be responding to changes in the halo mass function. Using a fixed total number of halos does not remove this sensitivity. The hypothesis was that the kNN-CDF signature is related to the different effects on the number of high and low mass halos, rather than the total number, and the different clustering of high and low mass halos. This differential effect would not be removed by using a fixed number of objects, as considered in this appendix.

Thus while the quantitative details are different, the results of this appendix qualitatively match the case where we use all the objects in the catalog.

\section{Convergence Tests}\label{app:convergence}
\begin{figure}
  \centering
  \includegraphics[width=.47\textwidth]{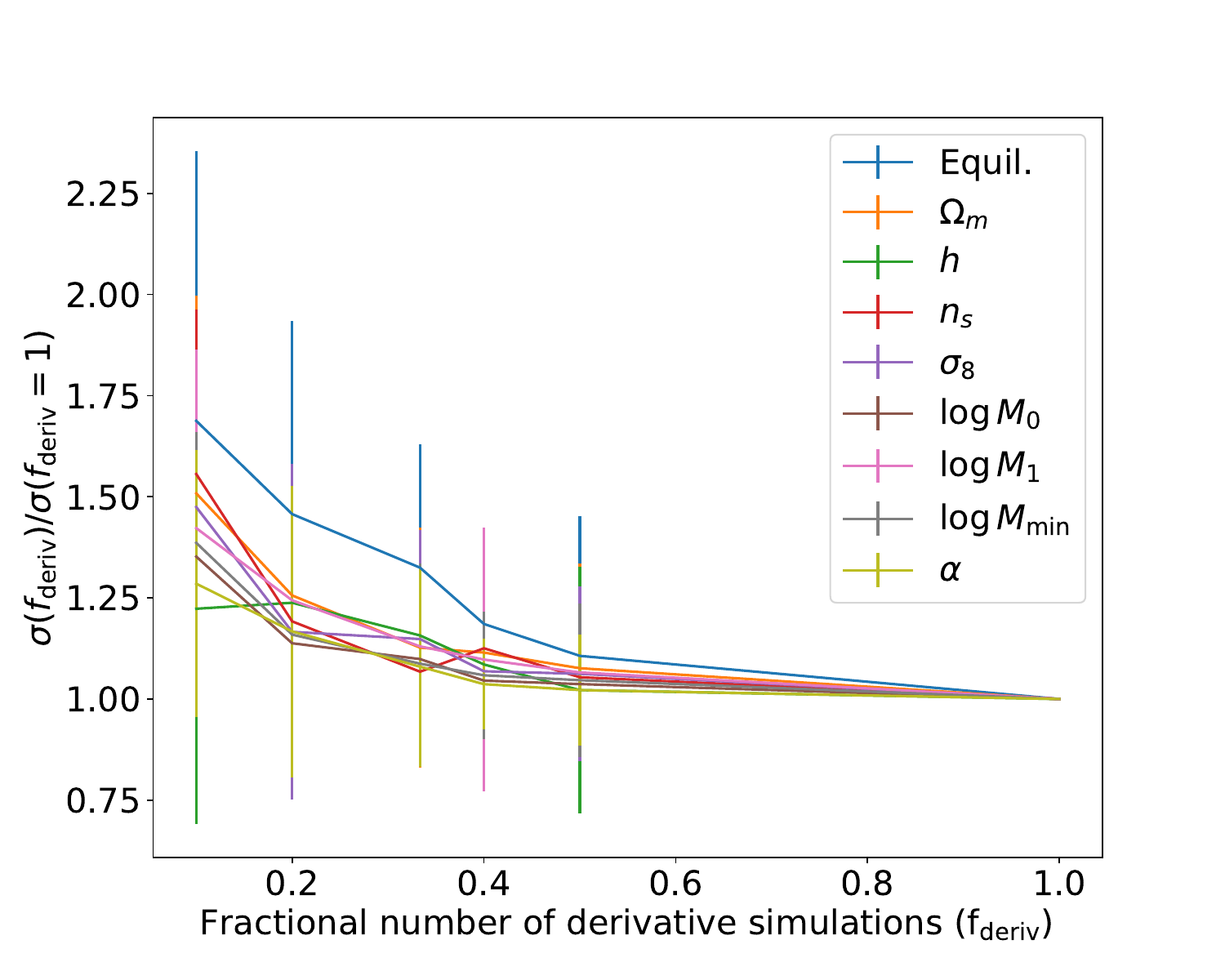} %height=.48\textwidth
\caption{ The compressed Fisher forecast parameter errors as a function of the fraction of the total number simulations used in the analysis. The error bars are estimated by the measured variances across different data spilts. See \citet{Coulton_2023b} for a more detailed discussion of compressed Fisher forecast convergences.
\label{fig:constraints_vs_nderivs}
}
\end{figure}

For simulation-based Fisher forecasts it is vital to test that the results are converged. Unconverged results arise from the noise associated with using a finite number of Monte Carlo simulations. In \cref{fig:constraints_vs_nderivs} we test the convergence our our galaxy Fisher forecast, \cref{sec:param_constraints}. The slow change in the forecast constraints with the number of simulations used implies that the forecast is sufficiently converged and so reliable.
\bibliographystyle{mnras}
\bibliography{references,planck_bib}

\end{document}